\documentclass[crcready,onecolumn]{iosart2x}

\makeatletter
\let\numberlines@hook\relax
\makeatother

\usepackage{newtxtext,newtxmath}
\hypersetup{colorlinks=true,linkcolor=blue,citecolor=blue,urlcolor=blue}

\pubyear{0000}
\volume{0}
\firstpage{1}
\lastpage{1}

\begin{document}

\begin{frontmatter}

\title{Energy Resolution and Neutron Flux of the 4SEASONS Spectrometer Revisited}
\runningtitle{Proceedings of ICANS-XXIII}

\author[A]{\inits{R.}\fnms{Ryoichi} \snm{Kajimoto}\ead[label=eKajimoto]{ryoichi.kajimoto@j-parc.jp}%
\thanks{Corresponding author. \printead{eKajimoto}.}},
\author[A]{\inits{M.}\fnms{Mitsutaka} \snm{Nakamura}\ead[label=eNakamura]{mitsutaka.nakamura@j-parc.jp}},
\author[B]{\inits{K.}\fnms{Kazuki} \snm{Iida}\ead[label=eIida]{k\_iida@cross.or.jp}},
\author[B]{\inits{K.}\fnms{Kazuya} \snm{Kamazawa}\ead[label=eKamazawa]{k\_kamazawa@cross.or.jp}},
\author[B]{\inits{K.}\fnms{Kazuhiko} \snm{Ikeuchi}\ead[label=eIkeuchi]{k\_ikeuchi@cross.or.jp}},
\author[A]{\inits{Y.}\fnms{Yasuhiro} \snm{Inamura}\ead[label=eInamura]{yasuhiro.inamura@j-parc.jp}}
and
\author[B]{\inits{M.}\fnms{Motoyuki} \snm{Ishikado}\ead[label=eIshikado]{m\_ishikado@cross.or.jp}}
\runningauthor{R. Kajimoto et al.}
\address[A]{Materials and Life Science Division, J-PARC Center, \orgname{Japan Atomic Energy Agency},
Tokai, Ibaraki 319-1195, \cny{Japan}\printead[presep={\\}]{eKajimoto,eNakamura,eInamura}}
\address[B]{Neutron Science and Technology Center, \orgname{Comprehensive Research Organization for Science and Society},
Tokai, Ibaraki 319-1106, \cny{Japan}\printead[presep={\\}]{eIida,eKamazawa,eIkeuchi,eIshikado}}

\begin{abstract}
The elastic energy resolution, integrated intensity, and peak intensity
of the direct-geometry neutron chopper spectrometer 4SEASONS at Japan
Proton Accelerator Research Complex (J-PARC) were re-investigated. This
was done with respect to the incident energy and the rotation speed of
the Fermi chopper using incoherent scattering of vanadium and simple
analytical formulas. The model calculations reproduced the observed
values satisfactorily. The present work should be useful for estimating
in instrument performance in experiments.
\end{abstract}


\end{frontmatter}


\section{Introduction}

The time-of-flight direct-geometry chopper spectrometer is one of the
typical and powerful neutron scattering spectrometers to measure atomic
and magnetic dynamics in materials. This type of instrument uses a
rotating chopper to monochromatize the neutron beam incident on a
sample, and the energy and momentum transfers to the sample are
determined by analyzing the time-of-flight of neutrons and scattering
angle of the detector. One of the advantages of this type of instrument
is that the energy and momentum ranges and resolution can be flexibly
chosen by tuning the rotation phase and speed of the chopper. This high
flexibility, however, sometimes makes it difficult to find the best
experimental condition in terms of resolution and flux, because high
resolution generally results in low flux. Therefore, it is important to
understand the relationship between the resolution and flux for the
instrument quantitatively and prepare a convenient tool to estimate
their values as a function of the incident energy ($E_\mathrm{i}$) and
the rotation speed ($f$) of the monochromatizing chopper. Monte Carlo
simulation can easily model the whole instrument including advanced
optics, and would be the best method to precisely reproduce the
resolution and
flux~\cite{lefmannNN99,willendrupPhysicaB04,willendrupJNR14,willendrupJNR19,vickeryJPSJ13,granrothEPJWebConf15,linNIMA16,kajimotoAIPConf18}.
However, Monte Carlo simulation takes too long time to use to decide the
experimental condition before or during the experiment, though it is
useful for data analysis after the experiment. To estimate the
experimental condition, a fast analytical calculation based on a more
simple model should be useful, even though it may be less accurate
compared with Monte Carlo simulations~\cite{linPhysicaB19}.

In this study, we investigated the energy resolution and flux of the
direct-geometry chopper spectrometer 4SEASONS at the Japan Proton
Accelerator Research Complex (J-PARC)~\cite{kajimotoJPSJ11}. 4SEASONS,
also called SIKI, is one of the four direct-geometry chopper
spectrometers installed at the pulsed neutron source of the Materials
and Life Science Experimental Facility (MLF) in
J-PARC~\cite{nakajimaQuBS17,kajimotoPhysicaB19}. It was designed for the
studies of dynamics using thermal neutrons, and has been used in a
variety of research fields such as superconductors, quantum magnets,
topological materials, catalysts, and thermoelectric materials. The
instrument views the supercritical hydrogen coupled moderator, which is
18\,m upstream of the sample. A Fermi chopper, 1.7\,m upstream of the
sample, is used to monochromatize the incident beam, and $E_\mathrm{i}$
of 10 to 250\,meV was typically used. Although the resolution and flux
have been analytically and numerically
investigated~\cite{kajimotoJPSJ11,iidaJPSConf14,kajimotoAIPConf18}, the
systematic investigation has not sufficiently been done, especially for
the flux. In addition, the Fermi chopper was replaced with a new model
in 2015, which has a short slit package consisting of 0.4-mm wide and
20-mm long slits~\cite{kajimotoJPCS18}. The new chopper was designed to
provide the same energy resolution as that of the old
chopper~\cite{nakamuraJPSConf14}, but no systematic studies of the
resolution and flux with the new chopper have been reported.

Accordingly, we re-investigated the energy resolution and intensity of
the elastic scattering using incoherent scattering of vanadium.  The
integrated intensity of the elastic scattering peak is the value
directly related to the neutron flux on the sample. On the other hand,
if the excitation to be observed is intrinsically sharp, the peak
intensity, rather than the integrated intensity, is essential to
determine the data quality. Therefore, in this study, we investigated
the integrated intensity and the peak intensity. By comparing the
observed data with simple analytical model, we developed empirical
formulas which are useful to calculate the energy resolution and
intensities.

\section{Experiment}

To study the scattering intensity as a function of $E_\mathrm{i}$ and
$f$, we measured scattering intensity of a vanadium sample while
rotating the Fermi chopper to monochromatize the incident beam. The
facility beam power was 0.51\,MW. The vanadium sample has a hollow
cylinder shape whose dimension is 18\,mm in diameter, 25\,mm in height,
and 1\,mm in thickness. The Fermi chopper was rotated at speeds in the
range of 100 to 600\,Hz with 100\,Hz steps. Additionally, to evaluate
the $E_\mathrm{i}$ dependence of the neutron flux without the chopper,
we measured scattering intensity of vanadium with a white beam. For the
latter measurement, we used a thin vanadium hollow cylinder whose
dimension is 20\,mm in diameter, 20\,mm in height, and 0.125\,mm in
thickness. For both measurements, the T0 chopper was rotated at 25\,Hz
to suppress background noise caused by the prompt pulse. The two disk
choppers were rotated at 25\,Hz to suppress the frame overlap for the
monochromatic beam measurement~\cite{kajimotoJPSJ11}. The scattering
intensity with the white beam and monochromatic beam were converted to
histogram data of neutron energy and energy transfer, respectively. To
obtain the peak width, also called full-width at half-maximum (FWHM),
and the peak height of the elastic peak in the monochromatic beam
measurements, we fitted the observed energy spectra to
Gaussians. Although the Gaussian fit sufficiently reproduced the
observed spectra peak widths and peak heights, the integrated
intensities of the obtained Gaussians underestimated the true integrated
intensities due to a pulse tail in the energy gain side, which is
particularly significant for instruments at the coupled moderator. Then,
we numerically integrated the observed energy spectra to obtain the
integrated intensities. The integrated intensity was converted to the
neutron flux per MW at the sample according to
Eq.~(\ref{eq_scatt_intensity}). The same conversion factor was applied
to the peak intensity per unit energy transfer (meV), although the
spectra as a function of energy transfer and their peak heights are
originally defined at the detector.

%

\section{Calculations}

\subsection{Energy resolution}




The energy resolution (FWHM) of the energy spectrum relative to the
incident energy ($E_\mathrm{i}$) for a Fermi chopper spectrometer is
described by the following
formula~\cite{iidaJPSConf14,windsor,ehlersRSI11,abernathyRSI12}:
\begin{equation}
 \label{eq_resolution}
 \frac{\Delta E}{E_\mathrm{i}} = \sqrt{
                 \left\{
                  2\frac{\Delta t_\mathrm{c}}{t_\mathrm{c}}
                  \left[1+\frac{L_1}{L_2}
                       \left(1-\frac{E}{E_\mathrm{i}}\right)^{\!\frac{3}{2}}
                  \right]
                 \right\}^{\!2}
		 +
		 \left\{
		  2\frac{\Delta t_\mathrm{m}}{t_\mathrm{c}}
		  \left[1+\frac{L_3}{L_2}
		       \left(1-\frac{E}{E_\mathrm{i}}\right)^{\!\frac{3}{2}}
		  \right]
		 \right\}^{\!2}
		 +
		 \left[
		  2\frac{\Delta L_2}{L_2}
		  \left(1-\frac{E}{E_\mathrm{i}}\right)\right]^{2}
	 },
\end{equation}
where $E$ is the energy transfer, and in this study
$E\!=\!\mathrm{0}$. The moderator-to-sample, sample-to-detector, and
chopper-to-sample distances are given as $L_1$, $L_2$, and $L_3$,
respectively. For 4SEASONS, $L_1\!=\!\mathrm{18\,m}$,
$L_2\!=\!\mathrm{2.5\,m}$, and $L_3\!=\!\mathrm{1.7\,m}$.  The time at
which neutrons with energy $E_\mathrm{i}$ reach the Fermi chopper is
given as $t_\mathrm{c}$. Similarly, the opening time of the Fermi
chopper and the pulse width at the moderator are denoted by $\Delta
t_\mathrm{c}$ and $\Delta t_\mathrm{m}$, respectively. Due to the
angular divergence of the incident beam, $\Delta t_\mathrm{c}$ is
effectively larger than its intrinsic value defined by chopper geometry,
i.e., $\Delta t_\mathrm{c} = p[w/(2{\pi}Df)]$, where $D$ and $w$ are the
diameter of the chopper rotor and width of each slit, respectively. The
term $p$ is expressed as a function of the maximum angular divergence of
the incident beam, $\Delta \Phi_i^\mathrm{max}$, as:
\begin{equation}
 p(u) = 
  \begin{cases}
   1+u/4 & 0 < u < 0.8 \\
   2+u-(4u-u^2)^{1/2} & 0.8 < u < 2 \\
   u & 2 < u,
  \end{cases}
\end{equation}
where
$u=\Delta\Phi_i^\mathrm{max}/(w/D)$~\cite{windsor}. $\Delta\Phi_i^\mathrm{max}$
has neutron energy dependence originating from neutron reflections by
the supermirrors of the guide tube. The $\Delta\Phi_i^\mathrm{max}$ was
estimated using the relationship between neutron wavelength and
supermirror critical angle~\cite{kajimotoJPSJ11,ikeuchiJPSJ13}. To
obtain $\Delta t_\mathrm{m}$, the linear interpolation of the numerical
values available at the J-PARC web-site~\cite{pulsedata} was
used. Furthermore, $\Delta L_2$ is the uncertainty of $L_2$ resulting
from the sample and the detector sizes. We assumed $\Delta L_2 = [(\pi
w_\mathrm{s}/4)^2 + (\pi w_\mathrm{d}/4)^2]^{1/2}$, where $w_\mathrm{s}$
and $w_\mathrm{d}$ are diameters of the sample and the detector,
respectively. $w_\mathrm{d}$ is 19\,mm for 4SEASONS, and 20\,mm for
$w_\mathrm{s}$ of the vanadium samples was used.

\subsection{Intensity}

The neutron flux at the sample is described as
follows~\cite{windsor,itohNIMA11}:
\begin{equation}
 \label{eq_flux}
 n(E_\mathrm{i},f) = \phi(E_\mathrm{i})\frac{S_\mathrm{m}}{L_1^2}G(E_\mathrm{i})
                     \frac{2(2/m)^\frac{1}{2}E_\mathrm{i}^\frac{3}{2}}{L_1-L_3}
                     \Delta t_\mathrm{c}
                     \frac{w}{w+z}A_\mathrm{c}(E_\mathrm{i}) T(E_\mathrm{i},f),
\end{equation}
where $\phi(E_\mathrm{i})$ is the neutron flux at the moderator,
$S_\mathrm{m}$ is the area of the moderator, $G(E_\mathrm{i})$ is the
gain by the guide tube, and $m$ is the neutron mass. The term $z$ is the
thickness of each neutron absorber in the slit package, and
$A_\mathrm{c}(E_\mathrm{i})$ is the absorption by the aluminum spacers
in the slit package. The transmission function $T(E_\mathrm{i},f)$ is
described as follows:
\begin{equation}
 T(\beta) =
  \begin{cases}
   1-\frac{8}{3}\beta^2 & 0< \beta< 1/4 \\
   \frac{16}{3}\beta^{1/2}-8\beta+\frac{8}{3}\beta^2 & 1/4 <\beta < 1 \\
   0 & \beta > 1,
  \end{cases}
\end{equation}
where $\beta = (D/2w)(\pi D f /v_\mathrm{i})$ for a straight slit Fermi chopper, and $v_\mathrm{i}$ is the neutron velocity.

%
%
%
%

Now we are interested in the $E_\mathrm{i}$ and $f$ dependence of the
scattering intensity detected by the detector,
\begin{equation}
 \label{eq_scatt_intensity}
 I(E_\mathrm{i},f) = n(E_\mathrm{i},f)N_\mathrm{v}\frac{\sigma_\mathrm{v}}{4\pi} A_\mathrm{s}(E_\mathrm{i}) \Delta\Omega\,\eta_\mathrm{d}(E_\mathrm{i}),
\end{equation}
where $\Delta\Omega$ is the solid angle, and $\eta_\mathrm{d}$ is the
efficiency of the detector. $N_\mathrm{v}$, $\sigma_\mathrm{v}$, and
$A_\mathrm{s}$ are the number of atoms in the vanadium sample, the
incoherent scattering cross-section of vanadium, and the absorption by
the vanadium sample, respectively.  We simplified the equation by
replacing terms independent of the Fermi chopper in Eq.~(\ref{eq_flux})
with the observed intensity of white beam,
$I_\mathrm{white}^\mathrm{obs}(E_\mathrm{i})$, and constant terms with a
scale factor, $C_0$. Then, we have the following formula for scattering
intensity as a function of $E_\mathrm{i}$ and $f$:
\begin{equation}
\label{eq_intensity}
 I(E_\mathrm{i},f) = C_0 I_\mathrm{white}^\mathrm{obs}(E_\mathrm{i})
                     E_\mathrm{i}^\frac{3}{2} \Delta t_\mathrm{c}
                     A_\mathrm{c}(E_\mathrm{i}) T(E_\mathrm{i},f).
\end{equation}
It should be noted that the intensity derived using this formula
corresponds to the integrated intensity. We calculated the peak
intensity using the energy resolution as shown in the results and
discussions section.

\section{Results and Discussions}

\begin{figure}[t]
 \includegraphics[scale=0.65]{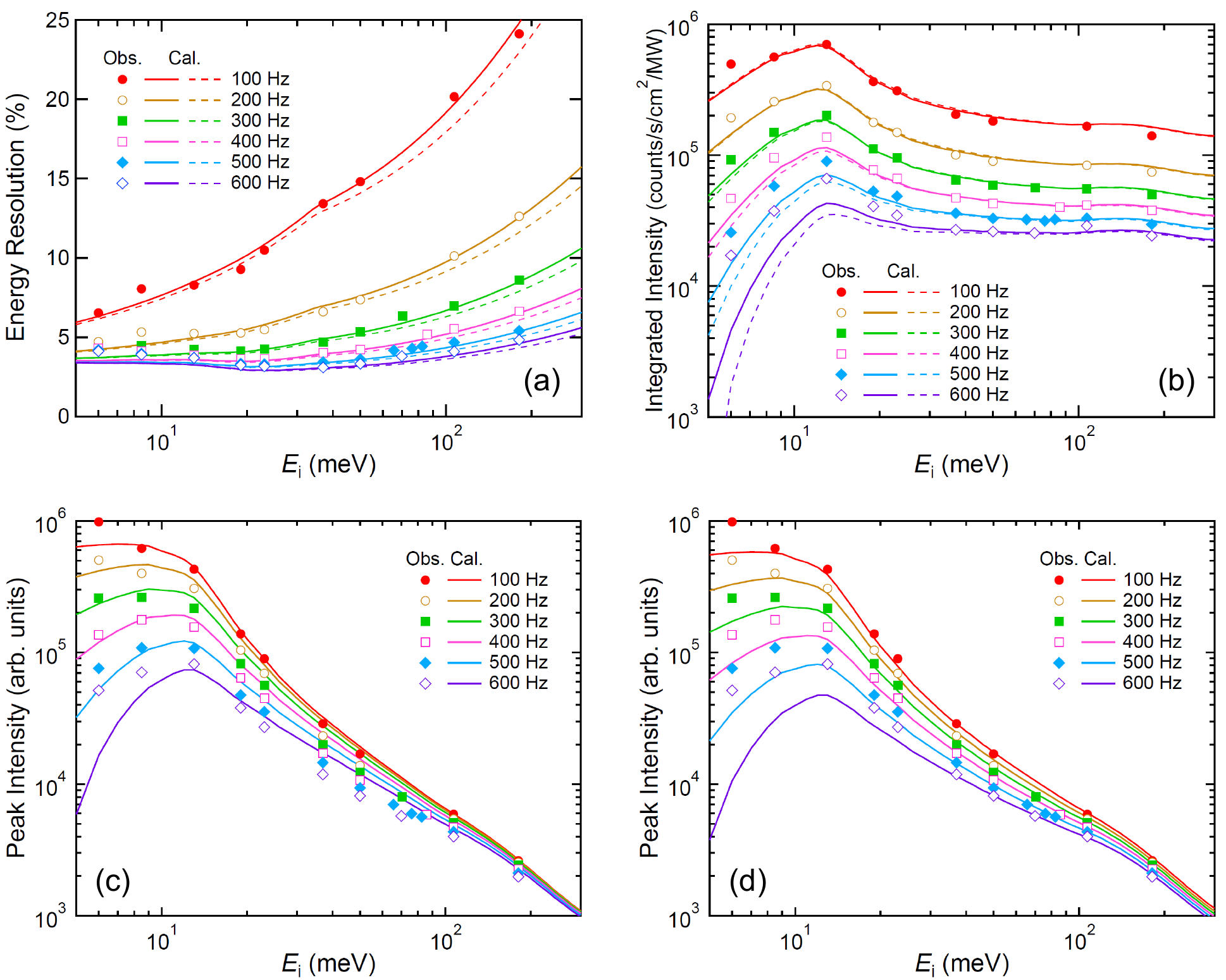}
 \caption{(a) The energy resolution, (b) integrated intensity, and (c,
 d) peak intensity at 4SEASONS as a function of the incident neutron
 energy for different Fermi chopper rotation speeds. Symbols denote the
 observed values. The values of the integrated intensities in (b) were
 converted to the neutron flux per MW at the sample. In (a) and (b), the
 broken lines and solid lines are calculated values with $(D,
 w)\!=\!(\mathrm{20\,mm, 0.4\,mm})$ and $(D,w)\!=\!(\mathrm{20\,mm,
 0.44\,mm})$, respectively. The solid lines in (c) and those in (d) are
 calculated values by using the formulas $I_\mathrm{peak} =
 0.73 (I/\Delta E)$ and Eq.~(\ref{eq_peak}),
 respectively.}
 \label{fig_Eres_intensity_amplitude}
\end{figure}

The $E_\mathrm{i}$ dependence of the observed values of the energy
resolutions and integrated intensities for several rotation speeds of
the Fermi chopper are shown in
Figs.~\ref{fig_Eres_intensity_amplitude}(a) and
\ref{fig_Eres_intensity_amplitude}(b), respectively. We found that the
intensities at $E_\mathrm{i}\!=\!\mathrm{6\,meV}$ were affected by the
finite transmission of the disk choppers around this energy region, and
the values were corrected for the transmission in
Fig.~\ref{fig_Eres_intensity_amplitude}(b). The broken lines in the
figures indicate the calculated values using Eqs.~(\ref{eq_resolution})
and (\ref{eq_intensity}), and the designed values of the geometrical
parameters of the Fermi chopper, i.e., $D\!=\!\mathrm{20\,mm}$ and
$w\!=\!\mathrm{0.4\,mm}$. The calculated integrated intensities were
normalized to coincide with the observed value at
$E_\mathrm{i}\!=\!\mathrm{19\,meV}$ and $f\!=\!\mathrm{300\,Hz}$. There
is good agreement between the observed and calculated values for the
energy resolutions and integrated intensities. However, the calculated
values are systematically lower than the observed values. Then, we
calculated the energy resolutions and the integrated intensities using a
10\% larger $w$ value, i.e., $w\!=\!\mathrm{0.44\,mm}$, which are
indicated by solid lines. Using this modification, the agreement with
the observed values improved for the energy resolutions and integrated
intensities. The larger slit width of the Fermi chopper may be due to
the fact that the absorber in the slit package is not 100\% neutron
absorber. The absorber of the 4SEASONS Fermi chopper comprised a blend
of 50\% $^{10}$B and 50\% epoxy glue. The latter may scatter neutrons on
the surface of the absorber, which can effectively widen the slit
width. We considered $w\!=\!\mathrm{0.44\,mm}$ as the practical slit
width, and used it for the other calculations hereafter.

As the rotation speed of the Fermi chopper increased, the energy
resolution improved
[Fig.~\ref{fig_Eres_intensity_amplitude}(a)]. However, it became
saturated as $f$ increased, as shown later in detail. The value of
energy resolution decreased as $E_\mathrm{i}$ is reduced, but slightly
increased below $\sim$20\,meV for high $f$s, because of the increase in
$\Delta t_\mathrm{m}$ of the coupled moderator. The energy dependence of
the integrated intensity showed a peak around 12\,meV, which corresponds
to the peak in the $\phi(E_\mathrm{i})E_\mathrm{i}^{3/2}$ term
[Fig.~\ref{fig_Eres_intensity_amplitude}(b)]. At high energies, the
integrated intensity decreased and became saturated as $f$ was
increased. In addition, a slight decrease in the calculated curves above
$\sim$150\,meV corresponds to the intensity loss due to the
slowly-rotating T0 chopper. On the other hand, at low energies, the
integrated intensity decreased quickly as $E_\mathrm{i}$ was
reduced. The intensity dropped more quickly at high $f$s because of the
rapid decrease in the transmission of the Fermi chopper,
$T(E_\mathrm{i},f)$. Nevertheless, the observed intensities at high
$f$s, especially at 500\,Hz and 600\,Hz, were clearly larger than the
calculated values. This tendency is consistent with the fact that the
observed amplitude of the energy resolution is larger than the
calculated one at these rotation speeds and low $E_\mathrm{i}$s.

\begin{figure}[t]
 \includegraphics[scale=0.65]{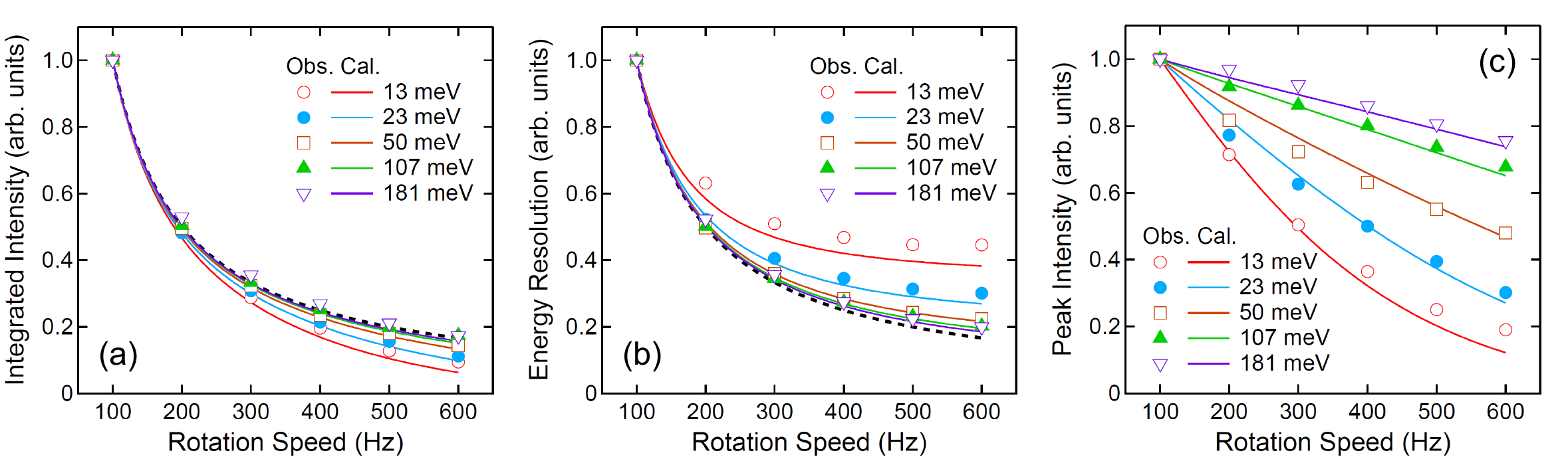}
 \caption{(a) The integrated intensity, (b) energy resolution, and (c)
 peak intensity at 4SEASONS as a function of the rotation speed of the
 Fermi chopper. Symbols denote the observed values, while solid lines
 denote the values calculated by using Eqs.~(\ref{eq_resolution}),
 (\ref{eq_intensity}), and (\ref{eq_peak}). The observed and calculated
 values are normalized so that the values at 100\,Hz are unities. In (a)
 and (b), the function of $100/f$ is drawn by the broken lines.}
 \label{fig_freq_dep}
\end{figure}

The symbols shown in Figs.~\ref{fig_Eres_intensity_amplitude}(c) and
~\ref{fig_Eres_intensity_amplitude}(d) show the observed peak
intensities as a function of $E_\mathrm{i}$. If the line shape of the
observed energy spectrum is a Gaussian, the peak intensity is
proportional to the integrated intensity divided by the peak
width. Thus, we calculated the peak intensities by $I_\mathrm{peak} =
C_1 (I/\Delta E)$, which are represented by the solid lines as shown in
Fig.~\ref{fig_Eres_intensity_amplitude}(c). The scale factor $C_1$ was
chosen to make the calculated intensity at 100\,Hz roughly reproduce the
observed values. However, this derivation was too naive, because the
calculated values deviated as $E_\mathrm{i}$ decreased and $f$
increased. One of the plausible reasons for the deviation was the fact
that the observed spectra were not Gaussians and had tails, which
became more significant at low $E_\mathrm{i}$s and high $f$s. Then, we
introduced the following empirical scale function which reduced the peak
intensity at low $E_\mathrm{i}$s and high $f$s:
\begin{equation}
 \label{eq_peak}
 I_\mathrm{peak} = \frac{I}{\Delta E} \times C_2   
  \left[ 1 + af^\frac{1}{2} \left(
			      b + \frac{1}{\exp{[c(E_\mathrm{i}-d)]}+1}
			     \right)
  \right]^{-1}
\end{equation}
A combination of parameters of $C_2\!=\!\mathrm{0.85}$,
$a\!=\!\mathrm{0.023}$, $b\!=\!\mathrm{0.5}$, $c\!=\!\mathrm{0.05}$, and
$d\!=\!\mathrm{70\,meV}$ resulted in the solid lines shown in
Fig.~\ref{fig_Eres_intensity_amplitude}(d). The agreement between the
calculated and observed values was improved compared with
Fig.~\ref{fig_Eres_intensity_amplitude}(c), and is similar to that for
the integrated intensity shown in
Fig.~\ref{fig_Eres_intensity_amplitude}(b). A better formula of the peak
intensity would be obtained if we could model the line shape precisely
by using an asymmetric function like the Ikeda-Carpenter
function~\cite{ikedaNIMA85}, because Eq.~(\ref{eq_peak}) and the
parameters listed above were empirically obtained. However, the
empirical formula of the peak intensity can be of practical use to
estimate the peak intensity as shown in
Figs.~\ref{fig_Eres_intensity_amplitude}(d) and \ref{fig_freq_dep}(c).


The observed and calculated energy resolutions, integrated intensities,
and peak intensities for selected $E_\mathrm{i}$s as a function of $f$
are shown in Fig.~\ref{fig_freq_dep}. The symbols denote the observed
values, while the solid lines represent values calculated by
Eqs.~(\ref{eq_resolution}), (\ref{eq_intensity}), and
(\ref{eq_peak}). The values were normalized so that those at 100\,Hz
were unity. The calculated values described the observed $f$ dependence
accurately in all the three kinds of data. In
Figs.~\ref{fig_freq_dep}(a) and \ref{fig_freq_dep}(b), the integrated
intensity and the energy resolution above $\sim$50\,meV followed a
function of $1/f$ (broken lines). This means that the $f$ dependences of
the integrated intensity and the energy resolution were dominated by
that of $\Delta t_\mathrm{c}$. However, they gradually deviated from the
$1/f$ dependence below $\sim$50\,meV. The deviation in the integrated
intensity resulted from the decrease in the transmission of the Fermi
chopper. However, the deviation was moderate, and the simple $1/f$ law
was an adequate measure to estimate the $f$ dependence of the integrated
intensity. On the other hand, the energy resolution significantly
deviated the $1/f$ dependence at low $E_\mathrm{i}$s, and it became
almost independent of $f$ at 13\,meV. This feature of the energy
resolution is correlated with the $f$ dependence of the peak intensity
as shown in Fig.~\ref{fig_freq_dep}(c). The peak intensity decreased as
a function of $f$. Although it showed moderate decrease as a function of
$f$ at high $E_\mathrm{i}$s, it decreased faster below $\sim$50\,meV,
where the energy resolution became saturated at high $f$s. The relation
between the integrated intensities, energy resolutions, and peak
intensities shown in Fig.~\ref{fig_freq_dep} indicated that at high
$E_\mathrm{i}$s above $\sim$50\,meV, the decrease in the integrated
intensity as a function of $f$ mostly resulted from the decrease in
$\Delta E$ while it was dominated by the decrease in the peak intensity
at low $E_\mathrm{i}$s below $\sim$50\,meV. This result suggests an
important criterion for choosing experimental conditions at 4SEASONS. In
general, high intensity and high resolution are mutually
exclusive. However, at high $E_\mathrm{i}$s above $\sim$50\,meV, the
decrease in the peak intensity as a function of $f$ was not as
significant as in the integrated intensity. Therefore, we can apply a
condition of high resolution with high peak intensity by rotating the
Fermi chopper at high speed. Such a condition is useful to measure
excitations which are sharp in energy such as crystal electric field
excitations and dispersionless optical modes. However, at low
$E_\mathrm{i}$s below $\sim$50\,meV, high speed rotation of the Fermi
chopper may waste both integrated and peak intensities without
significant improvement in the energy resolution.

\begin{figure}[t]
 \includegraphics[scale=0.65]{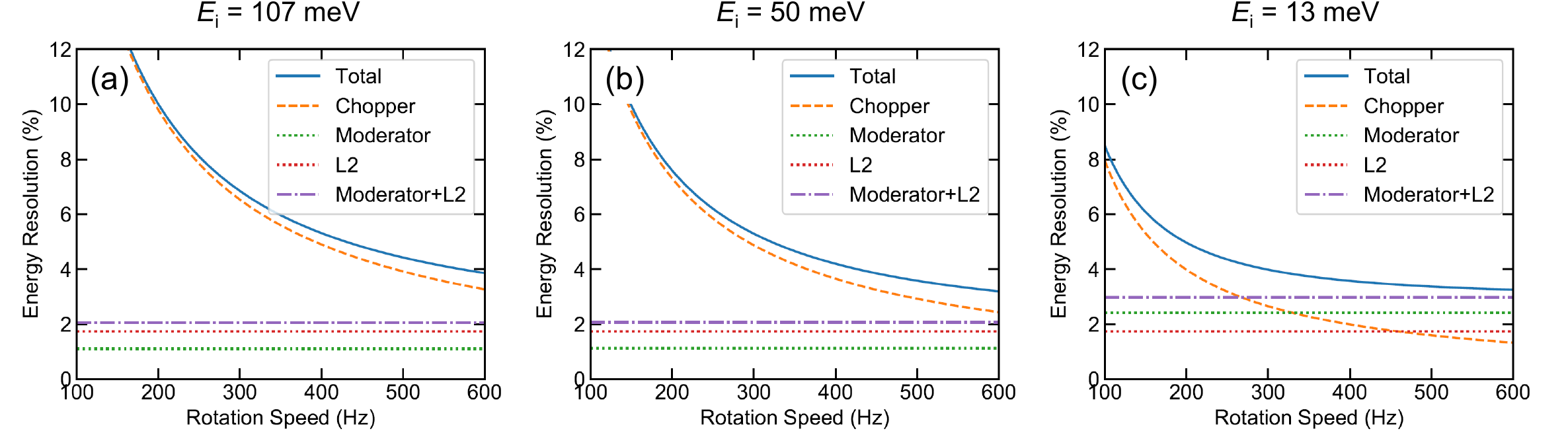}
 \caption{The calculated energy resolutions of 4SEASONS as a function of
 the rotation speed of the Fermi chopper for (a)
 $E_\mathrm{i}\!=\!\mathrm{107\,meV}$, (b) 50\,meV, and (c) 13\,meV. The
 solid lines denote the energy resolutions. The broken lines, green
 dotted lines, and red dotted lines denote the resolution components
 related to the chopper opening time ($\Delta t_\mathrm{c}$), moderator
 pulse width ($\Delta t_\mathrm{m}$), and the uncertainty in the
 sample-detector distance ($\Delta L_2$), respectively. The
 dashed-dotted lines denote the $f$-independent resolution components,
 which are square roots of sum of squares of the moderator and $L_2$
 components.}
 \label{fig_Eresolution}
\end{figure}

As seen above, the saturation of the energy resolution at low
$E_\mathrm{i}$s and high $f$s is an important feature in considering the
experimental condition of 4SEASONS. To understand this feature, we
plotted the calculated energy resolutions for
$E_\mathrm{i}\!=\!\mathrm{107}$, 50, and 13\,meV as a function of $f$ by
dividing them into respective resolution components in
Eq.~(\ref{eq_resolution}) as shown in Fig.~\ref{fig_Eresolution}. The
resolution component which originated from $\Delta t_\mathrm{m}$ [the
second term in Eq.~(\ref{eq_resolution})] and the resolution component
which originated from $\Delta L_2$ [the third term in
Eq.~(\ref{eq_resolution})] were constant regardless of the rotation
speed of the chopper. Typically, the $\Delta L_2$ component remains
constant for all $E_\mathrm{i}$s. It should be noted that the $\Delta
t_\mathrm{m}$ component was independent of $E_\mathrm{i}$ between
50\,meV and 107\,meV. This resulted from the fact that $\Delta
t_\mathrm{m}$ is proportional to $E_\mathrm{i}^{-1/2}$ in the epithermal
region~\cite{mildnerANE79}. However, the moderator component
significantly increased at $E_\mathrm{i}\!=\!\mathrm{13\,meV}$ because
of the increase in $\Delta t_\mathrm{m}$ of the coupled
moderator~\cite{pulsedata,takadaQuBS17}. On the other hand, the
resolution component which originated from $\Delta t_\mathrm{c}$ [the
first term in Eq.~(\ref{eq_resolution})] gradually decreased as
$E_\mathrm{i}$ decreased. As a result, at high energies such as
107\,meV, the energy resolution is mostly determined by the rotation
speed of the Fermi chopper
[Fig.~\ref{fig_Eresolution}(a)]~\cite{araiJPSj13}. Therefore, the energy
resolution improved at faster chopper rotation. At the middle energies
such as 50\,meV, the chopper component in the resolution was comparable
to the $f$-independent component (square root of sum of squares of the
moderator and $L_2$ components) at high $f$s
[Fig.~\ref{fig_Eresolution}(b)]. Finally, at low energies such as
13\,meV, the energy resolution was dominated by the $f$-independent
component for a wide range of $f$ [Fig.~\ref{fig_Eresolution}(c)], and
the energy resolution was not affected by $f$.

\begin{figure}[t]
 \includegraphics[scale=0.6]{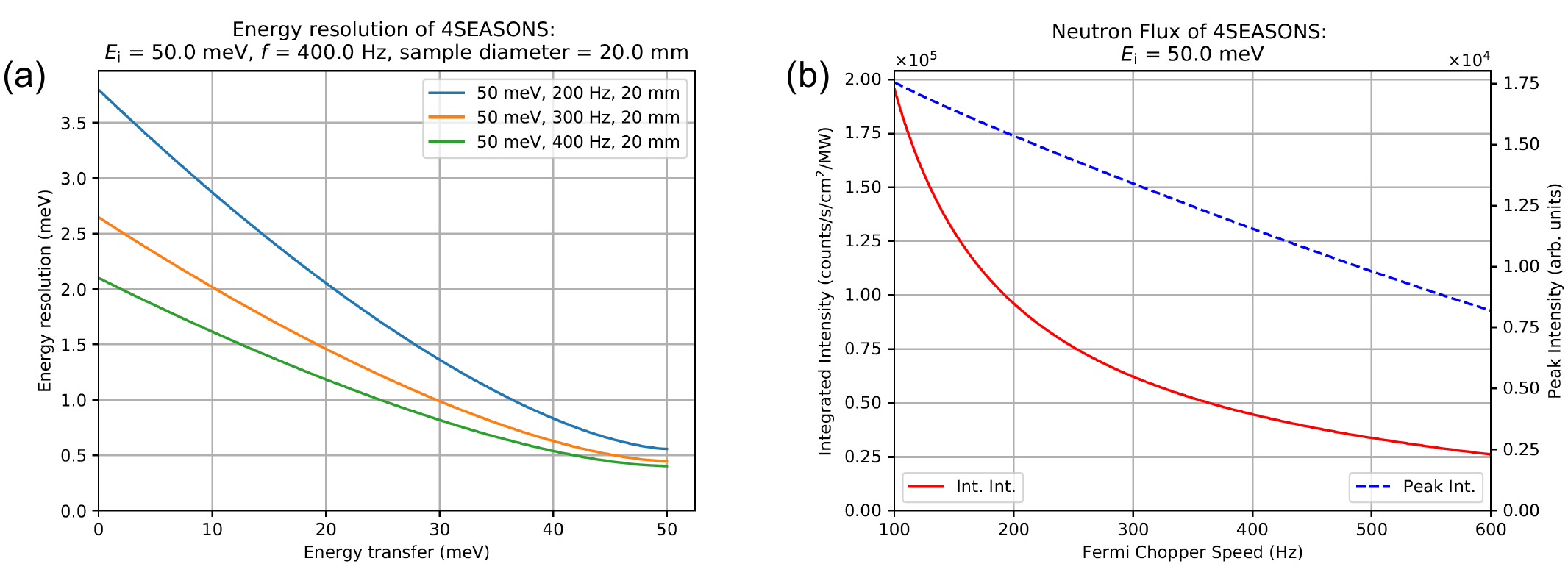}
 \caption{Outputs of computing tools to calculate (a) the energy
 resolution and (b) the integrated and peak intensities for 4SEASONS.}
 \label{fig_tools}
\end{figure}

Having established practical formulas to calculate the energy
resolution, integrated intensity, and peak intensity easily, we can take
advantage of this knowledge to find the best combination of
$E_\mathrm{i}$ and $f$ for experiments. For this purpose, we developed
simple scripts using Python programming language. An example of a script
output for the energy resolution is shown in
Fig.~\ref{fig_tools}(a). This script calculated the energy resolution as
a function of the energy transfer using Eq.~(\ref{eq_resolution}), and
has already been used at 4SEASONS. A prototype script output to
calculate the $f$ dependence of the integrated and peak intensities
recently developed based on the present study is shown in
Fig.~\ref{fig_tools}(b). These tools should be useful for users of the
instrument to choose the experimental conditions before or during
experiments.

\section{Summary}

The elastic energy resolution as well as the integrated and peak
intensities of the direct-geometry neutron chopper spectrometer 4SEASONS
at J-PARC were re-investigated with respect to the incident energy and
the rotation speed of the Fermi chopper using incoherent scattering of
vanadium. The model calculation based on formulas sufficiently
reproduced the observed energy resolution and integrated intensity, but
the agreement was improved by assuming a 10\% larger slit width in the
Fermi chopper. The naive division of the integrated intensity by the
peak width failed to reproduce the observed peak intensity probably due
to the asymmetric line shapes of the energy spectra, which became
significant at low incident energies and high rotation speeds. We
introduced an empirical scaling function, which practically described
the observed peak intensity. The inverse of the rotation speed of the
Fermi chopper is an adequate measure of the rotation speed dependence of
the energy resolution and integrated intensity at high incident
energies. However, they deviated from the $1/f$ law at low incident
energies as the transmission of the Fermi chopper decreased and the
resolution components which are independent of the chopper dominated the
energy resolution. Based on this study, simple computing tools to
calculate the energy resolution and intensities were developed, which
should be useful for users of the instrument to estimate the instrument
performance and decide the experimental condition before or during the
experiments.

\section*{Acknowledgment}

The neutron scattering experiment at MLF of J-PARC was performed under
the user program no.\ 2019I0001.



\bibliographystyle{ios1}           
\bibliography{icans23_kajimoto_200317_final}    

\end{document}